# A new soft X-ray magnetic circular dichroism facility at the BSRF beamline 4B7B[*]

GUO Zhi-Ying(郭志英)[1]  HONG Cai-Hao(洪才浩)[1]  XING Hai-Ying(邢海英)[2]  TANG Kun(唐坤)[1]  ZHENG Lei(郑雷)[1]

XUI Wei(徐伟)[1]  CHEN Dong-liang(陈栋梁)[1]  CUI Ming-Qi(崔明启)[1]  ZHAO YI-Dong(赵屹东)[1;1)]

1 Institute of High Energy Physics, Chinese Academy of Sciences, Beijing 100049, China

2 School of Electronics and Information Engineering, Tianjin Polytechnic University, Tianjin 300387, China

**Abstract:** X-ray magnetic circular dichroism (XMCD) has become an important and powerful tool because it allows the study of material properties in combination with elemental specificity, chemical state specificity, and magnetic specificity. A new soft X-ray magnetic circular dichroism apparatus has been developed at the Beijing Synchrotron Radiation Facility (BSRF). The apparatus combines three experimental conditions: ultra-high-vacuum environment, moderate magnetic fields and in-situ sample preparation to measure the absorption signal. We designed a C type dipole electromagnet that provides magnetic fields up to 0.5T in parallel (or anti-parallel) direction relative to the incoming X-ray beam. The performances of the electromagnet are measured and the results show good agreement with the simulation ones. Following film grown in situ by evaporation methods, XMCD measurements are performed. Combined polarization corrections, the magnetic moments of the Fe and Co films determined by sum rules are consistent with other theoretical predictions and experimental measurements.

## 1. Introduction

Spintronic materials [1, 2] such as diluted magnetic semiconductors (DMSs)[3], multiferroics [4], and Heusler alloys [5] have received much attention in science from the viewpoints of both academic research and applications. For the purpose of understanding the structure and function of these materials, the magnetic properties and electronic structure should be investigated and characterized. Furthermore, in order to clarify the origin of ferromagnetism in DMSs, it is necessary to use an element-specific technique.

* Supported by National Natural Science Foundation of China (61204008)
1) E-mail: zhaoyd@ihep.ac.cn

X-ray magnetic circular dichroism (XMCD) has been shown to be a powerful tool because it can be used to measure the element-specific magnetic moments [6]. This technique measures difference in absorption of left- and right- circularly polarized X-ray by a magnetized sample [7]. Generally speaking, the dichroic signal is proportional to the dot product of circular polarization vector $\vec{P}$ and magnetism vector $\vec{M}$. The spin and orbital moments can be independently determined by applying the MCD sum rules [8][9].So this technique has been widely used to measure the magnetic moments of specific element in samples. While almost all measured XMCD spectra by electron yield method contain effects due to saturation and incomplete circular polarization (especially for bending beamline). The saturation effects can be eliminated by choosing a suitable geometry and correcting the experimentally obtained data by a factor [10][11], but it is inconvenient and difficult for accurate determination of the correction factor. In this paper, we presented a new setup based on electromagnet for soft X-ray magnetic circular dichroism. By using this setup, we try to eliminate the impact of the two effects mentioned above by combination the transmission method with polarization correction. Furthermore, the point by point field reversal method has also been established for improving the measurement accuracy of magnetic moments.

## 2. Acquisition and analysis of the polarized light

4B7B is a soft x-ray beamline with a bending magnet, which used a variable-included-angle Monk-Gillieson mounting monochromator with a varied-line-spacing plane grating covering the energy range of 50eV~1600eV. The average photon flux with an energy resolving power of 3000 is about $10^9$ photons/s (250mA, 2.5GeV) at the L edge of Ar. The resolving power of the beamline is better than 6500 at 400eV by nitrogen absorption spectrum measurements. The circularly polarized light can be obtained by positioning the beamline aperture out of the plane of the electron storage ring. The degree of the circularly polarized light at the end of the beamline was analyzed in our previous work [12]. A multilayer polarimeter [13] was used to investigate the effects of aperture position on polarization and intensity at the $L_{2,3}$ edges of 3d transition metals. And we find that 70% circular polarization can be obtained by moving the aperture to a position of half-maximum intensity.

## 3. XMCD experimental set up

In addition to the circularly polarized light, another necessary experimental condition of XMCD is the magnetic field to magnetize the sample. A variety of magnets have been developed and applied in XMCD instruments. In 1990, Chen and co-workers [14] used a permanent magnet to observe the soft XMCD effect in Ni. The permanent magnet should be reversed manually or mechanically to switch the field, but it is still used by some groups [15, 16]. Electromagnets are widely used by several groups in a variety of ways [17-22]. The main advantage of an electromagnet over a permanent magnet is that the direction and intensity of the magnetic field can be automatically manipulated. Superconducting magnets [23] [24] can produce greater magnetic fields than all, but it is expensive and the field ramping rates is lower than electromagnets for applying point-by-point field reversal method.

Based on above considerations, we have developed a dipole electromagnet that has the following characteristics: large magnetic gap for sufficient sample space, moderate magnetic field intensity for general magnetic materials, high symmetry and low remanence in sample position. Other aspects of the XMCD experiment are also considered and these considerations finally determined the design of the system.

Fig. 1 shows the main parts of the XMCD set-up, which consists of a UHV chamber constructed of 316L non-magnetic stainless steel in order to reduce the remanence. This rectangular chamber is fixed between the magnetic poles of an electromagnet. The electromagnet consists of two C-type magnets. Each C-type magnets is composed of coils of insulated copper wire wrapped around a low-carbon iron core. We used hollow copper conductor which has a square cross section(6mm by 6mm) with Ø4mm round hole for water cooling. A DC power supply with high current and low voltage (200A，20V) are used for the electromagnet. The maximum temperature rise in the electromagnet is about 18° at a current of 150A, which is consistent with our calculation result. The two magnets are joined together by stainless steel plates. The connecting plates have a central hole through which a 14mm tube joins the chamber to the beamline. The upper part of the set up is a XYZ vacuum manipulator for sample adjustment. The lower part is for sample in-situ preparation and pumping system. The entire devices were mounted together and positioned at the beamline focus. The chamber, which is sandwiched between the electromagnet, was carefully collimated to ensure that the whole light spot can pass through. The base pressure of the system is about $1\times10^{-7}$pa and can be improved to $1\times10^{-8}$pa after back out.

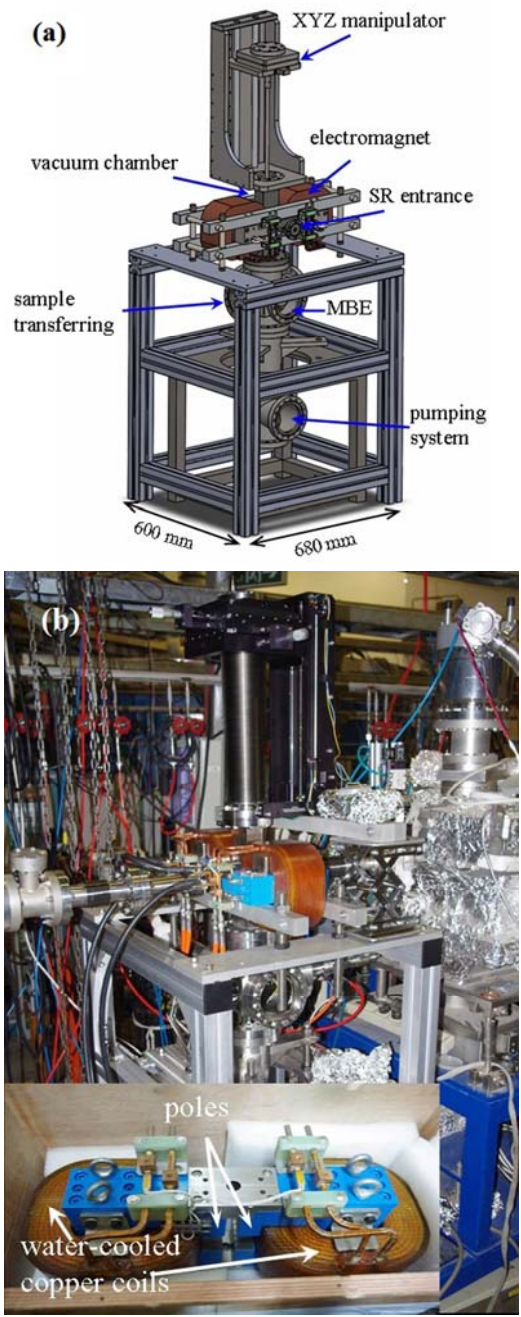

Fig.1. (a) The mechanical design of the XMCD apertures is shown in this diagram. (b) The photograph of the electromagnet (the inset below) and the entire system.

The magnet was designed with a large magnetic gap for sufficient sample space. The magnetic poles are separated by a gap of 45mm, with 40mm of internal space in the chamber. The geometry of the iron cores and the coils was optimized to obtain a homogeneous magnetic field in the sample region. The intensity and the distribution of the magnetic field are simulated by vector field software OPERA [25]. Fig.2 shows the three-dimensional magnetic field distribution analyzed by this software. Fig.3 (a) presents the calculated and measured current-field relationship at the sample region.The experimental results show good agreement with the model calculations. The remanent field at the sample region was less than 5Gs after applying 4400Gs and it is always less than 0.12% of the maximum field for a hysteresis loop measured. The calculated and measured magnetic field distribution are showed in Fig.3 (b), it can be found that the uniformity of the magnetic field is better than 3% within ±10mm.

The sample holder is designed to be compatible with two usual kinds of detection methods: TEY (total electron yield) and transmission method. The TEY mode is widely used for general sample in soft x-ray absorption spectroscopy. But when it is applied in XMCD, it contains saturation effects and the results need to be corrected. The transmission mode is not affected by this effect and is also suitable for insulating sample.

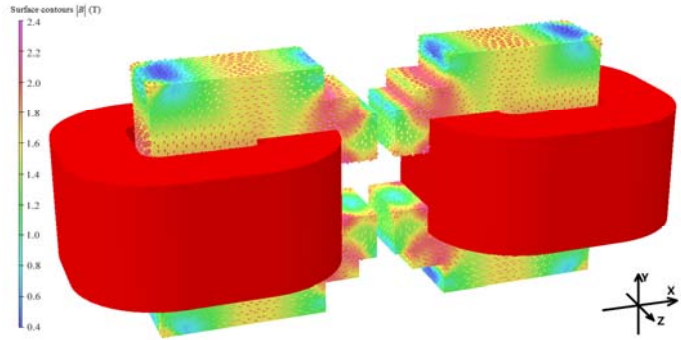

Fig.2. The three-dimensional magnetic field distribution with coil currents of 150 A.

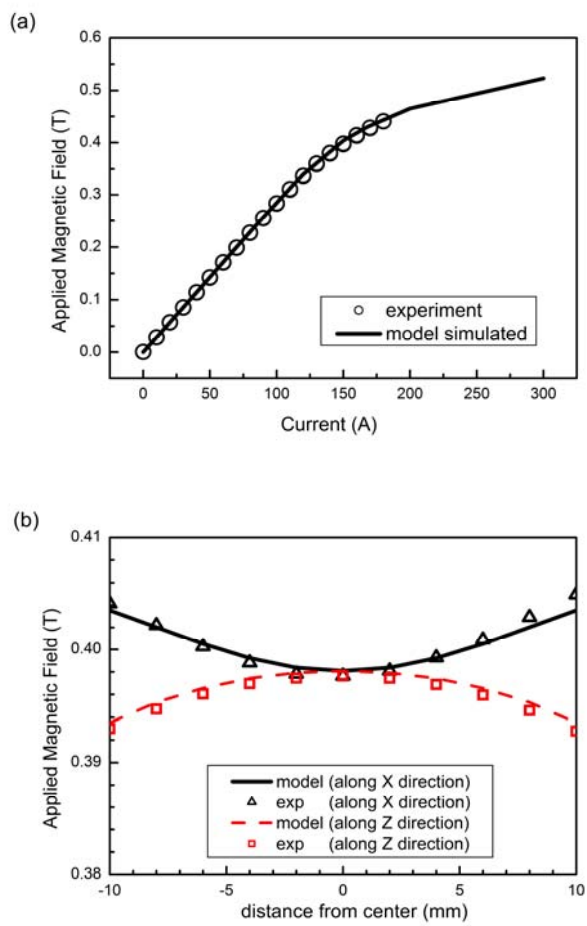

Fig.3. (a) Calculated and measured current-field relationship at the sample position. (b) Calculated and measured the distributions of the magnetic field for Y=0 plane at coil currents of 150 A.

## 4. Results and discuss

**4.1 XMCD from Co ultra-thin film by transmission method**

To verify the applicability of this XMCD set up and to eliminate the two effects mentioned above for determination the precise magnetic momentum, we have measured X-ray absorption spectroscopy (XAS) and XMCD spectra of Co ultra-thin film by the transmission method with polarization correction. The beam intensity before the sample ($I_0$) is monitored by the electron yield from the refocusing mirror. The intensity after the sample ($I$) is measured by a photodiode SX100 from IRD (International Radiation Detection Inc.). All current signals are detected by Model 6517B electrometer (Keithley Instruments Inc.). Co films were in situ grown by e-beam evaporation onto self-supported amorphous CH substrates with a transmission of about 90%. The evaporation rate was monitored by a quartz crystal rate/thickness monitor (SQM 160, INFICON, USA) and kept at 0.05Å/s for 12 minutes. The photon incident angle was set at 90° and the degree of circular polarization was 70% which was measured by the polarization analysis device.

XAS spectra across the $L_{2,3}$ edges of Co were measured at two opposite magnetization directions with the magnetic field of ±0.4T. Fig.4(a) are the raw data of $I_0$ and the transmission signal of Co thin films with two opposite magnetization directions. After background subtraction and normalization, the standard absorption spectra are shown in Fig.4(b). According to the XMCD sum rules [17], the orbital and spin moments can be determined by the following expressions:

$$m_{orb} = -\frac{2\mu_B}{3C}(A+B), \quad m_{spin} = \frac{\mu_B}{C}(-A+2B) \qquad (1)$$

where the A, B and C are the appropriate integration shown in Fig 4(c). In these expressions the $\langle T_z \rangle / \langle S_z \rangle$ term are neglected and $\mu_0$ are assumed to be equal to $(\mu^+ + \mu^-)/2$. The final results, corrected by the degree of circular polarization, are determined and listed in Table 1 to compare with other theoretical predictions and experimental measurements. The spin moments show good agreement with the gyromagnetic results. It proved that the spin moments do not need to be corrected by the $\langle T_z \rangle / \langle S_z \rangle$ term mentioned above. Unexpectedly, the orbit moments show small enhancement compared with others' results. This result can be explained by the low-dimensional structures which may be present on the surface of the sample and contribute the extra magnetic moments. And there are some other factors which will increase the uncertainties of the experimental results, for example the possible variations in the two-step function, the non-symmetric influence of the magnetic fields [29] and the discrepancies in the data processing. So these factors should be carefully considered for high-precision magnetic moment detection.

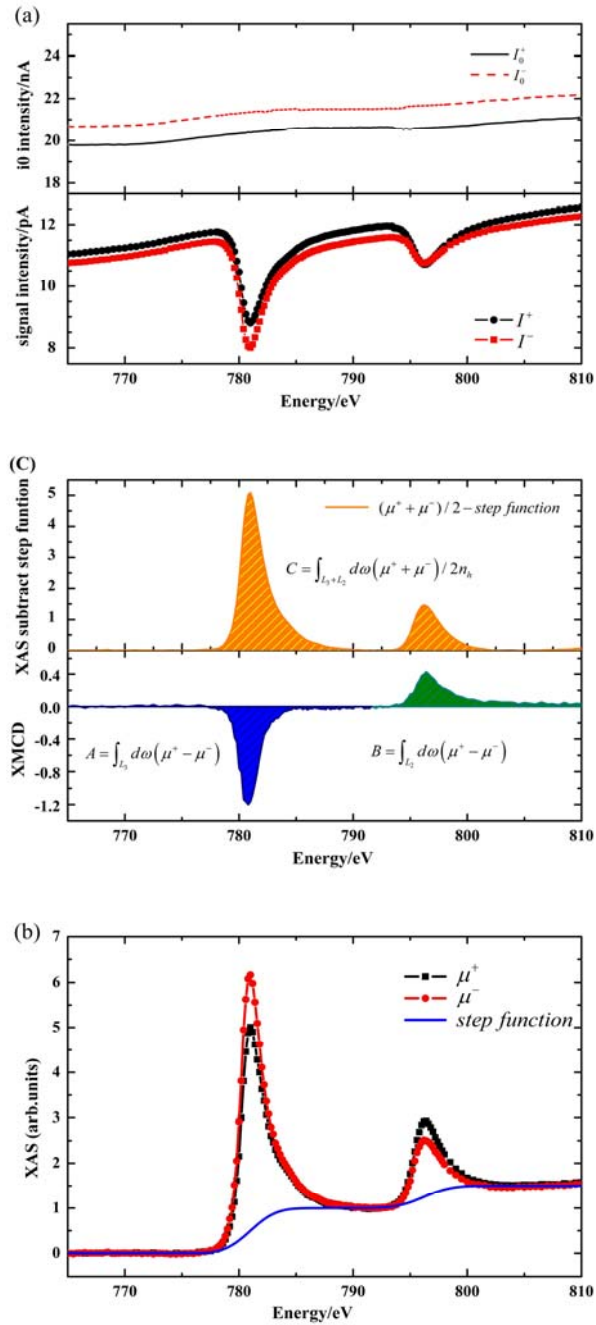

Fig.4. $L_{2,3}$-edge XAS and MCD spectra of cobalt: (a) The raw date of $I_0$ and the transmission signal of Co thin films with two opposite magnetization directions; (b) the standard absorption spectra calculated from the raw date, and the blue line(color figures on the web) is the two step function for edge-jump removal before the integration; (c) is the summed XAS spectra after removal the step function and the MCD spectra, the relationship between the parameters A,B,C and the integration are also shown beside the spectra.

Table 1. Spin and orbital moments for Co in units of $\mu_B$/atom.

|  |  | Co (hcp) | | | Fe (bcc) | | |
|---|---|---|---|---|---|---|---|
|  |  | $m_{orb}/m_{spin}$ | $m_{orb}$ | $m_{spin}$ | $m_{orb}/m_{spin}$ | $m_{orb}$ | $m_{spin}$ |
| Experiment | This work | 0.109 | 0.167 | 1.52 | 0.043 | 0.088 | 2.03 |
|  | C.T.Chen MCD[26] | 0.095 | 0.154 | 1.62 | 0.043 | 0.085 | 1.98 |
|  | C.T.Chen corrected $\langle T_z \rangle / \langle S_z \rangle$ | 0.099 | 0.153 | 1.55 | 0.043 | 0.086 | 1.98 |
|  | Gyromagnetic ratio | 0.097 | 0.147 | 1.52 | 0.044 | 0.092 | 2.08 |
| Theory | OP-LSDA [27,28] | 0.089 | 0.140 | 1.57 | 0.042 | 0.091 | 2.19 |

**4.2 XMCD from Fe film by TEY method**

Although the TEY method have the problem of saturation effects that make it is difficult to correct the experimental data, it is still the most popular method for XAS and XMCD experiment. And it is suitable for general samples. To test this method in the XMCD set-up, we performed in-situ measurements on Fe films grown on Al substrate by e-beam evaporation in the parasitic mode of BSRF. Fig.5 shows the absorption spectra for the $L_{2,3}$ edges of Fe measured with ±0.4T of applied magnetic field(150A) and 60% circularly polarized light. We used the values of 3.39 for the number of Fe 3d holes. The correction factor of saturation effects was estimated to be 0.75. After correction by the two factors (saturation effects and incomplete circular polarization), the spin and orbit moments determined by the sum rule are listed in Table 1 and show agreement with the gyromagnetic ratio measurements and the XMCD results by C.T.Chen.

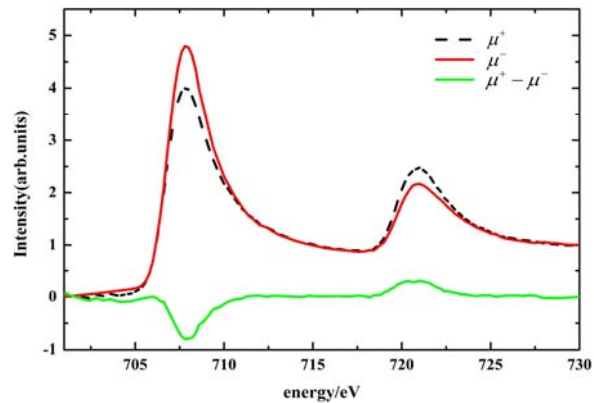

Fig.5. Fe $L_{2,3}$-edge spectra and MCD date measured at room temperature and in normal incidence by TEY method. The spectra in this graph have not been corrected by the degree of circular polarization and the correction factor for saturation effects.

**4.3 XMCD improved by each point field reversal method**

The XMCD setup can be operated in two modes: measuring twice for different field direction and point-by-point field reversal method. All above data are measured by the first mode, and the experimental results show that it can be easily affected by the orbit drift and the beam attenuation, and thus made it difficult for spectral normalization and spectral subtraction. So we established the each (energy) point field reversal method. The ramping rates can be adjusted by the power supply, and it generally takes about 1s for each point. This mode is time consuming for constant reverse the field but the results are more credible and easier to handle. This method can be easily used in the dedicated mode at BSRF which has a long beam-lifetime compare to the parasitic mode.

## 5. Conclusions

This paper reports on the performance of the new setup developed for XMCD measurements at the BSRF beamline 4B7B. This device, which combines three experimental conditions: ultra-high-vacuum environment, moderate magnetic fields and in-situ sample preparation, are proved to be suitable for XMCD measurements. We have shown that the high precise magnetic momentum can be determined by combination the transmission method with polarization correction. The values so determined show good agreement with others' theoretical predictions and experimental measurements. It was realized that the current experimental setup is limited to the ferromagnetic samples due to the room temperature conditions, so in the future we plan to add the sample cooling and temperature control units to the present system in order to establish the temperature-dependent XMCD. One possible application of this method is the origin of room temperature ferromagnetism in DMS. And it is possible to provide the experimental evidence for understanding the coupling mechanism at the interface of FM/DMS bilayer.

*The authors would like to thank the magnet group and the power supplies group for their help during the design and construction of the electromagnet.*